\documentclass[prb,aps,twocolumn,superscriptaddress]{revtex4-1}
\usepackage{graphicx,color}
\usepackage{amsthm}
\usepackage{amsfonts}
\usepackage{algorithmic}
\usepackage{enumerate}
\usepackage{latexsym}
\usepackage{amsmath}
\usepackage{amssymb}
\usepackage[colorlinks=true,citecolor=blue,linkcolor=blue]{hyperref}
\def\avg#1{\left\langle#1\right\rangle}

\emergencystretch=\maxdimen
\begin{document}

\title{Metal-insulator transition in the disordered Hubbard model of the Lieb lattice}

\author{Yueqi Li}
\affiliation{Department of Physics, Beijing Normal University, Beijing
100875, China}

\author{Lingyu Tian}
\affiliation{Department of Physics, Beijing Normal University, Beijing
100875, China}

\author{Tianxing Ma}
\email{txma@bnu.edu.cn}
\affiliation{Department of Physics, Beijing Normal University, Beijing
100875, China}

\author{Hai-Qing Lin}
\affiliation{Beijing Computational Science Research Center, Beijing
100193, China}
\affiliation{Department of Physics, Beijing Normal University, Beijing
100875, China}


\begin{abstract}
Using the determinant quantum Monte Carlo method, we investigate the metal-insulator transition in the interacting disordered Hubbard model of a Lieb lattice,
in which the system characterizes the flat band centered at the Fermi level.
By choosing suitable electron densities,
we ensure the weak interaction sign problem to improve the reliability of our results.
It is found that disorder and on-site Coulomb repulsive interaction produce interesting effects that induce the metal-insulator transition which is impossible in the half-filled case.
The density of states at the Fermi energy is still finite in the thermodynamic limit, suggesting that the system is an Anderson insulator rather than a Mott insulator.
Moreover, this doping system is paramagnetic, unlike the half-filled system, which is ferrimagnetic.

\end{abstract}
\maketitle

\section{\label{intro}Introduction}
A series of experiments have demonstrated the appearance of a conducting phase in
a Si/SiGe heterostructure\cite{PhysRevB.56.R12741}
and silicon metal-oxide-semiconductor
field-effect transistors
\cite{PhysRevB.51.7038,PhysRevLett.77.4938,RevModPhys.73.251}
as the electron density is varied.
These findings contradict the standard scaling theory of localization, which claims that
metal-insulator transition (MIT) cannot exist in a two-dimensional noninteracting system with disorder\cite{PhysRevLett.42.673}.
In a low-density system, since the ratio of
Coulomb repulsion to kinetic energy is large,
the electron-electron interaction is important in the disordered system
and is likely responsible for the metallic behavior.

The interplay of disorder and interaction has been widely studied in different physical
systems\cite{Vasseur2004,PhysRevB.65.125308,PhysRevB.81.075106} and has led to several interesting conclusions.
For example, using quantum Monte Carlo (QMC) approaches to the Anderson-Hubbard model of a square lattice,
repulsive interaction significantly
enhances the DC conductivity and then induces a metal-insulator transition at quarter filling\cite{PhysRevLett.83.4610}.
On the honeycomb lattice, the study found that disorder suppresses long-range anti-ferromagnetic order and
induces a novel nonmagnetic insulating phase\cite{PhysRevLett.120.116601}.
All of these phenomena illustrate the necessity of considering electronic correlation in disordered systems.

\begin{figure}
\flushleft
\begin{minipage}[c]{0.23\textwidth}
\centering
\includegraphics[width=3.5cm]{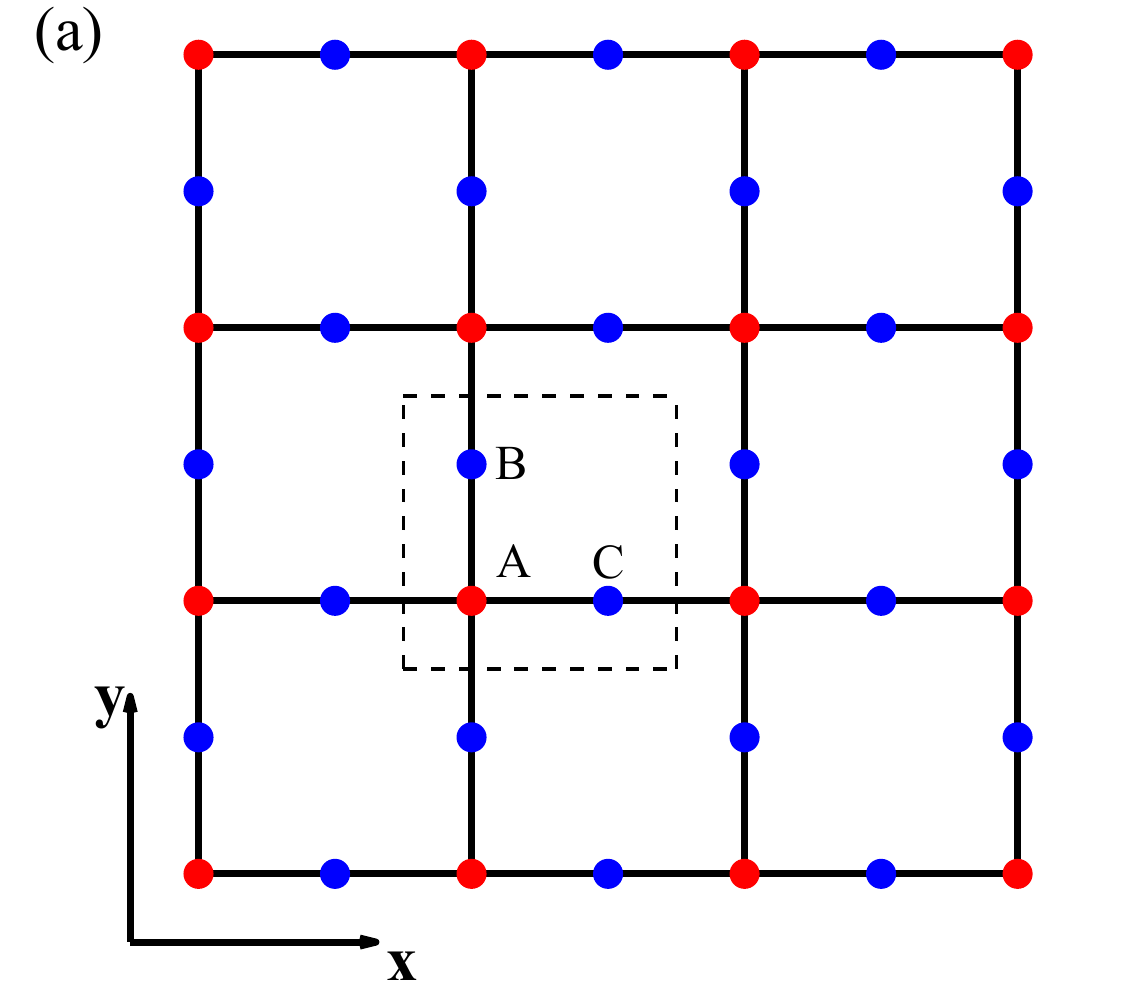}
\end{minipage}
\begin{minipage}[c]{0.2\textwidth}
\centering
\includegraphics[width=5cm]{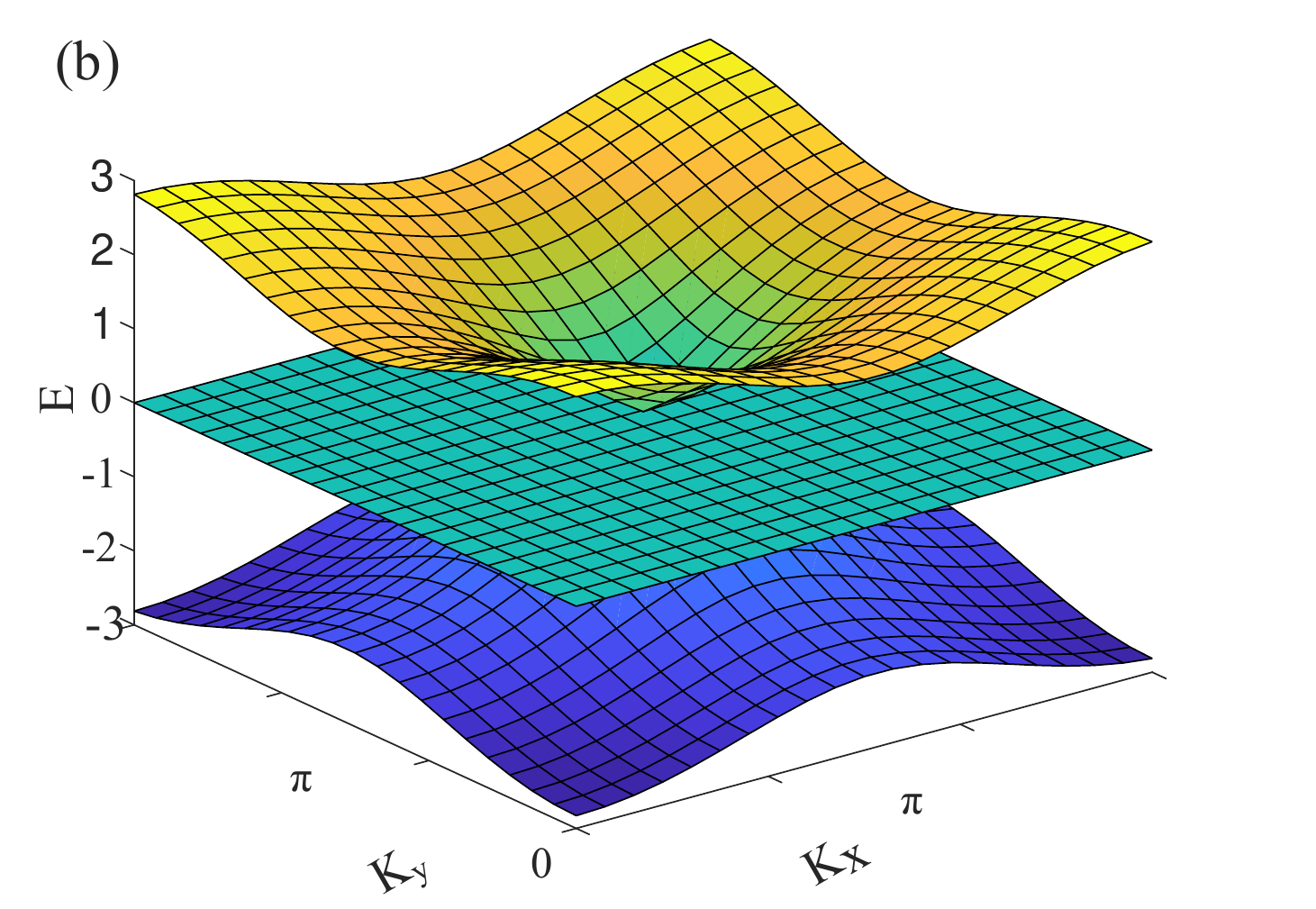}
\end{minipage}
\caption{(a) Structure of the Lieb lattice. The dotted square shows one unit cell containing three lattice sites
with indices A, B, and C.
(b) The band structure of the tight-binding model of the Lieb lattice.}
\label{Fig1}
\end{figure}

The flat band system, which ranges from optical lattices \cite{Taiee1500854,PhysRevLett.118.175301,PhysRevB.81.041410} to ultracold atomic
gases\cite{Hart2015,PhysRevX.8.031045} to photonic devices\cite{Zong:16,PhysRevB.85.205128,PhysRevLett.114.245503,PhysRevLett.114.245504},
is another issue in modern condensed matter physics.
Although it has been theoretically understood for three decades, this system continues to receive a great deal of attention,
in particular with regards to its ability to realize new many-body phases \cite{PhysRevLett.117.045303,PhysRevB.95.024515,Peotta2015,PhysRevLett.109.096404}.
Twisted bilayer graphene is a striking example of this\cite{Cao2018,Cao20182}.
One of the simplest flat band 2D systems is the Lieb lattice, which is also known
as the side-centered square lattice.
There are three types of sites, labeled A, B and C, in each unit cell outlined by the dotted square
in Fig.\ref{Fig1} (a).
For the infinite lattice with only nearest-neighbor hopping, the Lieb lattice is a three-band
structure that consists of two dispersion bands
$\epsilon_{\pm}(\bold{k})=\pm t\sqrt{4+2\cos(\bold{k}
\cdot \bold{a}_1)+2\cos(\bold{k} \cdot \bold{a}_2)}$ and one dispersionless, macroscopically degenerate
flat band $\epsilon(\bold{FB})=0$ centered at the Fermi level.
Tthese three bands touch each other at the $M$ point $(\pi,\pi)$, where the low energy
spectrum behaves as a single Dirac cone in the Brillouin zone, as shown in Fig.\ref{Fig1} (b).

Recently, the electronic Lieb lattice has been realized in surface state electrons using CO molecules and Cu(111),
providing a realistic system to explore physical properties\cite{Slot2017,Drost2017}.
Furthermore, in experiment, the atoms can also be arranged into the Lieb lattice, similar to the CuO$_{2}$ plane of cuprate superconductors.
\cite{PhysRevB.82.085310,Mielke_1991,PhysRevB.44.770,PhysRevLett.58.2794}.
A closer correlation between the theoretical Lieb structure physical materials gives this model more value.

The Hubbard model is the simplest model to use in investigation of the interaction between particles.
At half filling, studies of the flat band Lieb lattice have revealed interesting physics:
the system behaves as a topologically trivial gapless insulator when the Coulomb repulsion $U=0$,
but it behaves as a Mott insulator when $U>0$\cite{PhysRevB.96.245127},
and the topological phase transition caused by the spin-orbit coupling is connected to
the presence of a nondispersive flat band\cite{PhysRevB.100.045420}.
The Lieb theorem states that bipartite lattices with two sublattices that have different sites,
$N_{A}\neq N_{B}$, have a ferromagnetic state at half filling
with on-site Coulomb repulsion only when electron hopping between different sublattices is considered\cite{PhysRevLett.62.1201}.
The QMC simulations have shown that the local moment is strongly dependent on B/C sites due to decreased itinerancy caused by fewer neighbors,
and a metal without any magnetic ordering is found only if the repulsive effect is considered at A sites\cite{PhysRevB.94.155107}.
In addition, when the interaction strength at sites is switched off randomly, an insulator-to-metal transition and magnetic transition will occur simultaneously\cite{PhysRevB.101.165109}.
However, because of the infamous sign problem, most studies have focused on half filling, and research on doping is
highly important.

In this paper, our simulations are completed by the exact determinant quantum Monte Carlo (DQMC) method for cases with different fillings.
For systems with varying on-site Coulomb repulsion, $U$, disorder, $\Delta$, and electron density, $\avg{n}$, the DC conductivity is calculated to distinguish between metallic and insulating phases.
For conditions far from half filling, we focus on $\avg{n}=0.3$ and $0.5$, for which the sign problem is less significant compared to cases with other densities, and we reveal
the appearance of $U-$- and $\Delta-$-inducing metal-insulator transitions that are absent in the half-filled case.
We find that repulsion plays a delocalized role on electrons, whereas disorder plays the opposite role.
In the thermodynamic limit, the finite density of states at Fermi energy suggests that the insulator is an Anderson insulator instead of a Mott insulator.
Our results extend the understanding of the metal-insulator transition and magnetic order in the Lieb lattice.

\section{\label{intro}Modelling methods}

Our model is characterized by an interacting disordered Hubbard model of a Lieb lattice. The Hamiltonian, using periodic boundary
conditions, reads:
\begin{eqnarray}
H&=&H_{\textrm{0}}+H_{\textrm{$\mu$}}+H_{\textrm{U}} \notag \\
H_{\textrm{0}}&=&-\sum_{\alpha \alpha' \langle \bf{i,j} \rangle \sigma}  t_{\bf ij} (\hat c_{\alpha {\bf i} \sigma}^\dagger \hat c_{ \alpha' {\bf j} \sigma}^{\phantom{\dagger}}+\mbox{H.c.}) \notag \\
H_{\textrm{$\mu$}}&=&-\mu \sum_{\alpha \bf i \sigma}\hat c^{\dag}_{\alpha \bf i \sigma} \hat c_{\alpha \bf i \sigma}\notag \\
H_{\textrm{U}}&=&U \sum_{\alpha \bf i }(\hat n_{\alpha \bf i \uparrow}-\frac{1}{2})(\hat n_{\alpha \bf i \downarrow}-\frac{1}{2})
\end{eqnarray}
where $\langle \bf{i, j}\rangle$ runs over all pairs of nearest-neighbor sites on the lattice, and $\hat c_{\alpha { \bf i}\sigma}^\dagger(\hat c_{\alpha {\bf i}\sigma}^{\phantom{\dagger}})$ is the creation (annihilation)
operator for fermion on site $i$ of the $\alpha (\alpha')$ sublattice $(\alpha / \alpha'=A,B,C)$, satisfying anti-commutation
relations $\{\hat c_{\alpha {\bf i}}^{\dag}, \hat c_{\alpha {\bf j}}\}=\delta_{ij}$ and $\{\hat c_{\alpha {\bf i}}, \hat c_{\alpha {\bf j}}\}=0$, and H.c. stands for the Hermitian conjugate.
Disorder is introduced by taking the hopping parameters, $t_{\bf ij}$, from a probability distribution, $P(t_{\bf ij})=1/\Delta$, for
$t_{\bf ij}\in [t-\Delta/2,t+\Delta/2]$, and zero otherwise. $\Delta$ is a measure of the strength of the disorder, and $t$, the unit of energy,
is set to 1 throughout our paper. The on-site Coulomb repulsive interaction is parameterized by $U$
and the electronic densities of system $\avg{n}$ are controlled by chemical potential $\mu$.
When $\mu$=0, our method ensures that the system becomes half-filled and remains particle-hole symmetric
without the sign problem, even with the existence of bonding disorder.
$\hat n_{\alpha \bf i \sigma} = \hat c^{\dag}_{\alpha \bf i \sigma} \hat c_{\alpha \bf i \sigma}\notag$ is the number operator on site ${\bf i}$ of the $\alpha$ sublattice.

We employ the DQMC approach to study the Lieb lattice for on-site Coulomb repulsion.
In this method, the partition function, $Z$, is expressed as $Z=Tre^{-\beta H}$ and is then approximated by a path integral using Trotter-Suzuki decomposition\cite{10.1143/PTP.56.1454,PhysRevB.33.6271},
which divides the imaginary-time interval $[0,\beta]$ into $M$ equal subintervals of width $\Delta \tau=\beta/M$.
The interaction term formed in the term of quartic fermion operators is decoupled through a Hubbard-Stratonovich (HS) transformation\cite{PhysRevB.31.4403},
and thus introduces the fluctuating fields by Ising variables.
The resulting Hamiltonian quadratic form can be integrated analytically and represented as the product of
the determinant of spin up and spin down determinants.
The physical quantity proceeds with Monte Carlo sampling, which stochastically changes the space- and imaginary time-dependent
auxiliary field.
One DQMC sweep is composed of traversing the entire site, and in our simulation, we have used 8000 warmup sweeps
to arrive at equilibrium state and 24000 measuring sweeps to compute.
There are two sources of statistical error in our results: one is induced by the Trotter error $\Delta\tau$,
We set $\Delta\tau=0.1$, which confirms that the error generated is small enough to neglect\cite{PhysRevB.85.125127}; the other is due to
disorder averaging, and we reduce this error by averaging the results from $20$ disorder realizations. Ref.\cite{PhysRevB.104.045138} have proved
that the computed data are already consistent for realization numbers larger than 10.

The metal-insulator transition can be observed by the behavior of $T$-dependent DC conductivity, which can be obtained
using the wave vector ${\bf q}$ and the imaginary (Matsubara) time $\tau$-dependent current-current correlation function $\Lambda_{xx}$.
\begin{eqnarray}
\label{DC}
\sigma_{dc}(T)=\frac{\beta^2}{\pi}\Lambda_{xx}(\textbf{q}=0,\tau=\frac{\beta}{2}) ,
\end{eqnarray}
where $\Lambda_{xx}(\textbf{q},\tau)=<\hat{j}_x(\textbf{q},\tau)\hat{j}_x(\textbf{-q},0)>$,
$\beta$ is the reciprocal of temperature, and $\hat{j}_x(\textbf{q},\tau)$ is the Fourier transform of the
current density operator $\hat{j}_x(\textbf{i},\tau)$ in the $x$ direction.
This formula is valid only if the temperature is lower than the energy scale,
Thus, the high-order term of the Taylor expansion formula of the frequency-dependent function of conductance needs to be sufficiently small\cite{PhysRevB.101.245161}.
In the disordered system we study, the energy scale is set by the disorder strength $\Delta$ so that Equation (\ref{DC}) is valid at low temperatures.
The equation has been used to determine either a metallic
or an insulating phase in the disordered Hubbard model in many studies\cite{PhysRevLett.75.312,PhysRevLett.83.4610,PhysRevB.54.R3756}.
A further way to distinguish the type of insulating phase is the density of states (DOS) at Fermi energy\cite{PhysRevLett.75.312,Lederer4905}.
\begin{eqnarray}
\label{DOS}
N(0)\simeq \beta \times G({\bf r} = 0,\tau = \beta/2),
\end{eqnarray}
where $G$ is the imaginary time Green function.
We are also concerned about the magnetic behavior of the system by focusing on the space equal-time spin-spin correlation function\cite{PhysRevB.80.075116}, defined as
\begin{eqnarray}
\label{SS}
C(r)=<S^{z}(\textbf{R}_{i}+\textbf{r})S^{z}(\textbf{R}_{i})>,
\end{eqnarray}
where $S^{z}(\textbf{R}_{i})=n_{i\uparrow}-n_{i\downarrow}$.
From $C(\textbf{r})$, we can obtain spin structure factor $S(\textbf{q})$ through the Fourier transform:
\begin{eqnarray}
\label{Spin}
S(\textbf{q})=\sum_{\textbf{r}}e^{i\textbf{q}\cdot\textbf{r}}C(\textbf{r}).
\end{eqnarray}
Then, we define the $\textbf{q}$-dependent susceptibility as $\chi(\textbf{q}) =\beta S(\textbf{q})$\cite{PhysRevLett.83.4610}.
The wave vectors $\textbf{q}=(0,0)$ yield the ferromagnetic (FM) susceptibility, and$\textbf{q}=(\pi,\pi)$
yield anti-ferromagnetic (AF) susceptibility\cite{PhysRevB.84.155123}.
The AF structure factor in our paper is calculated by the formula $S_{AFM}=S^{z}_{AA}+S^{z}_{BB}+S^{z}_{CC}-S^{z}_{AB}-S^{z}_{AC}-S^{z}_{BA}-S^{z}_{CA}+S^{z}_{BC}+S^{z}_{CB}$,
where each term that contributes to AFM is considered, so AFM results are unaffected no matter how the electrons are distributed on the A/B/C sites\cite{GOUVEIA2016292}.

\section{\label{intro}Results and discussion}

\begin{figure}[t]
\centerline {\includegraphics*[width=3.6in]{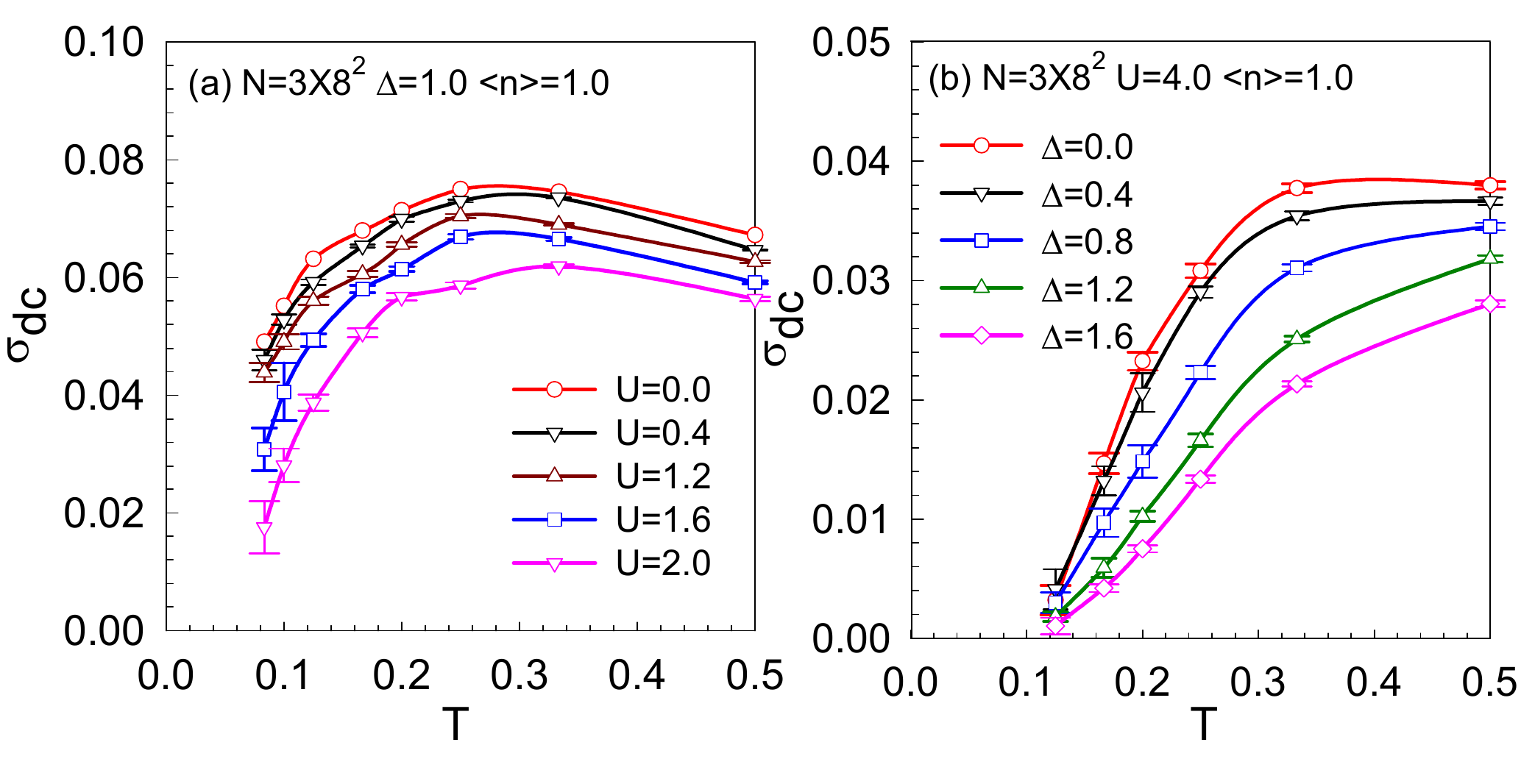}}
\caption{(Color online) Temperature dependence of the DC conductivity $\sigma_{dc}$ computed on the $N_{c}=3\times8\times8$
lattice for (a) various interaction strengths at $\Delta=$1.0 and (b) various disorder strengths at $U=4.0$. Data points are averages over $20$ disorder realizations.}
\label{Fig:half}
\end{figure}

First, we plot the DC conductivity as a function of temperature under a few representative $U$ and $\Delta$
at half filling.
Fig.\ref{Fig:half} (a) shows that keeping the disorder strength $\Delta=1.0$,
The DC conductivity decreases with increasing repulsive interaction strength.
In the case of higher temperatures, $\sigma_{dc}$ increases as the temperature decreases,
while in the case of $T\leq 0.2$, $\sigma_{dc}$ starts to decrease,
indicating insulating behavior.
The effect of disorder on the DC conductivity with the intermediate repulsive interaction $U=4.0$ is shown in Fig.\ref{Fig:half} (b),
where increasing the hopping disorder strength reduces the value of the DC conductivity.
Neither increases in the hopping disorder strength nor the interaction strength can change the insulating state at half filling.
Our finding that there is no metal-insulator transition in the hopping
disordered Hubbard model at half filling is consistent with previous results\cite{PhysRevB.94.155107}.

\begin{figure}[t]
\centerline {\includegraphics*[width=3.6in]{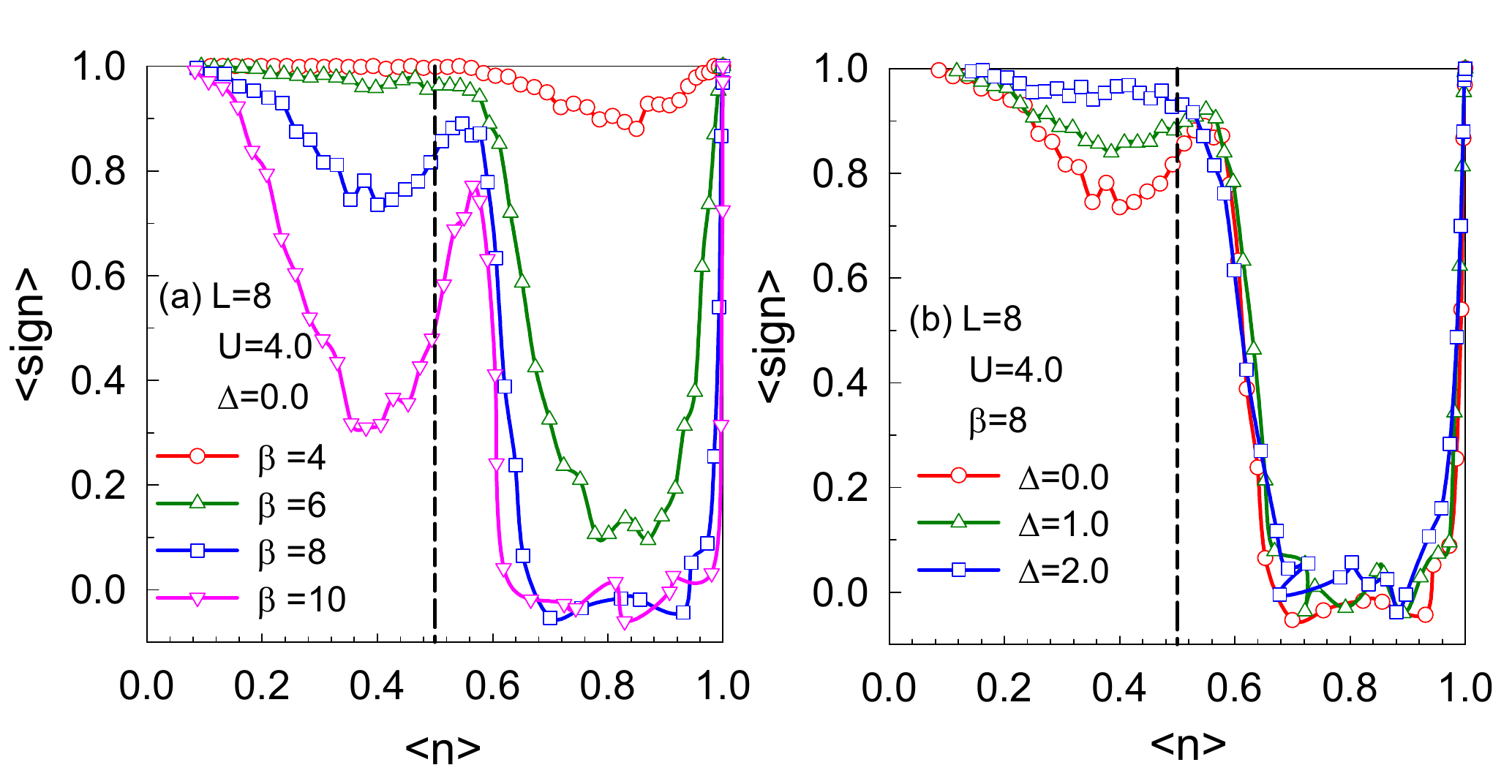}}
\caption{(Color online) Sign problem $\avg{sign}$ as a function of density for
(a) Different temperatures and (b) various disorder strengths. The dashed line represents $\avg{n}=0.5$.}
\label{Fig:sign}
\end{figure}

Our simulation is aimed at understanding the physical properties when the band is far from half filling.
The fermion sign problem is protected by the particle-hole symmetry at exactly half filling, while our simulations are plagued by the sign problem when doping away from half filling,
\cite{PhysRevB.42.2282}, and this restricts accessible temperatures.
To avoid the error caused by the sign problem, we compute the relation between the sign problem and density
for different temperatures and disorder strengths at fixed $L=8$ and $U=4.0$ in Fig. \ref{Fig:sign}.
As $\avg{n}$ decreases from $1$, the average fermion sign $\avg{sign}$ decreases rapidly to $0$
and floats about $0$ until $\avg{n}=0.7$.
Then, as $\avg{n}$ gradually decreases to $0$, $\avg{sign}$ recovers to $1$.
$\avg{sign}$ shows no signal as it passes through $\avg{n}=2/3$, which corresponds to entry into the flat band.
Fig.\ref{Fig:sign} (a) shows the inverse temperature $\beta$ dependence of the sign; increasing $\beta$ leads to a more severe sign problem.
Indeed, the average sign decays exponentially with increasing $\beta$\cite{PhysRevB.92.045110}.
Conversely, increasing the disorder strength ameliorates  the sign problem, as shown in Fig.\ref{Fig:sign} (b).
To ensure the reliability of the results, we choose two densities of $\avg{n}=0.3$ and $\avg{n}=0.5$ to study the effect of
doping on the transport and dynamic properties, and the results were obtained with more than $24000$ iterations at lower temperatures to reduce errors.

\begin{figure}[t]
\centerline {\includegraphics*[width=3.6in]{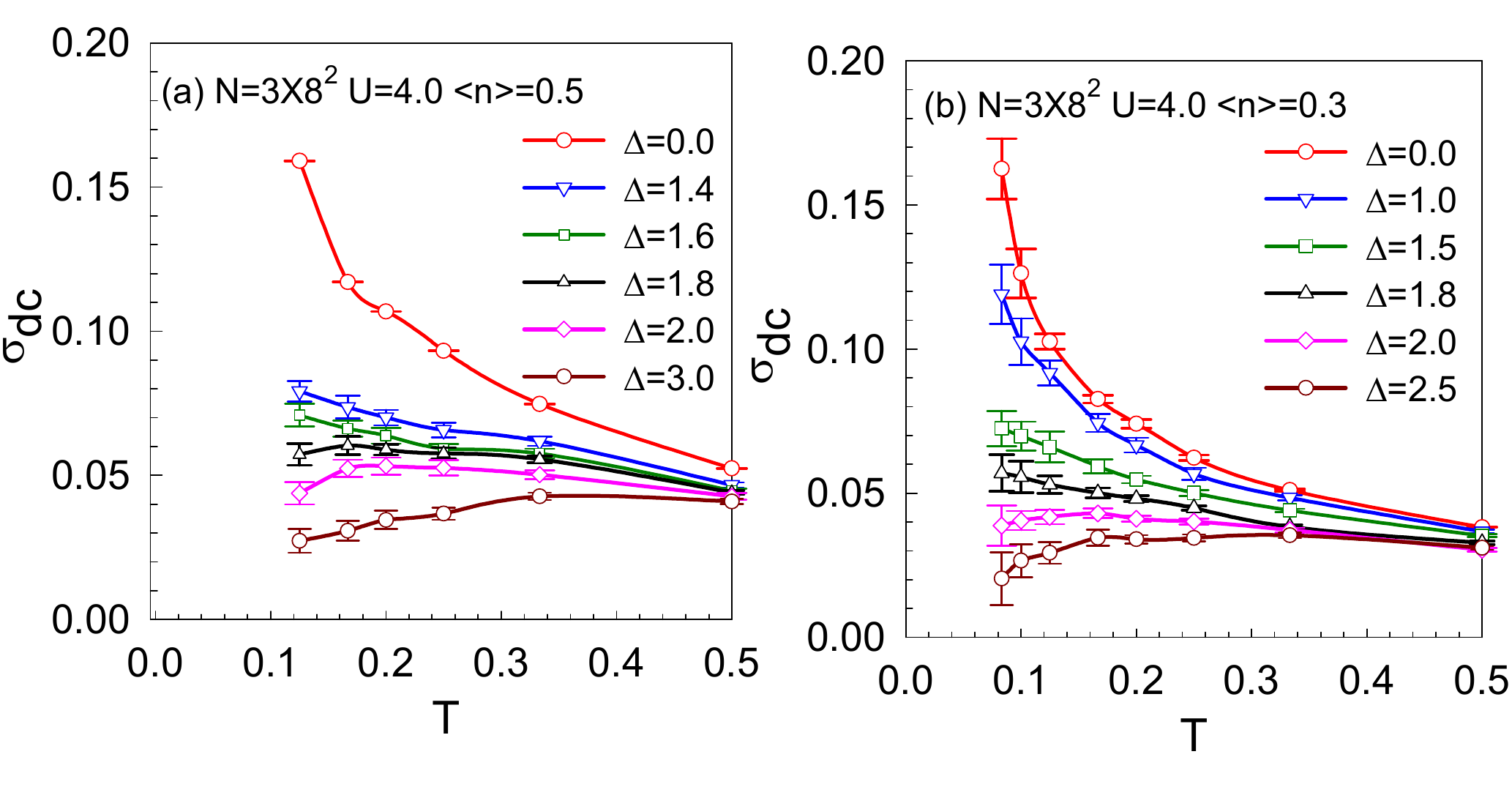}}
\caption{(Color online) The DC conductivity as a function of temperature for various disorder strengths at $\avg{n}=0.5$ and
$\avg{n}=0.3$.}
\label{Fig:dc}
\end{figure}

\begin{figure}[t]
\centerline {\includegraphics*[width=3.6in]{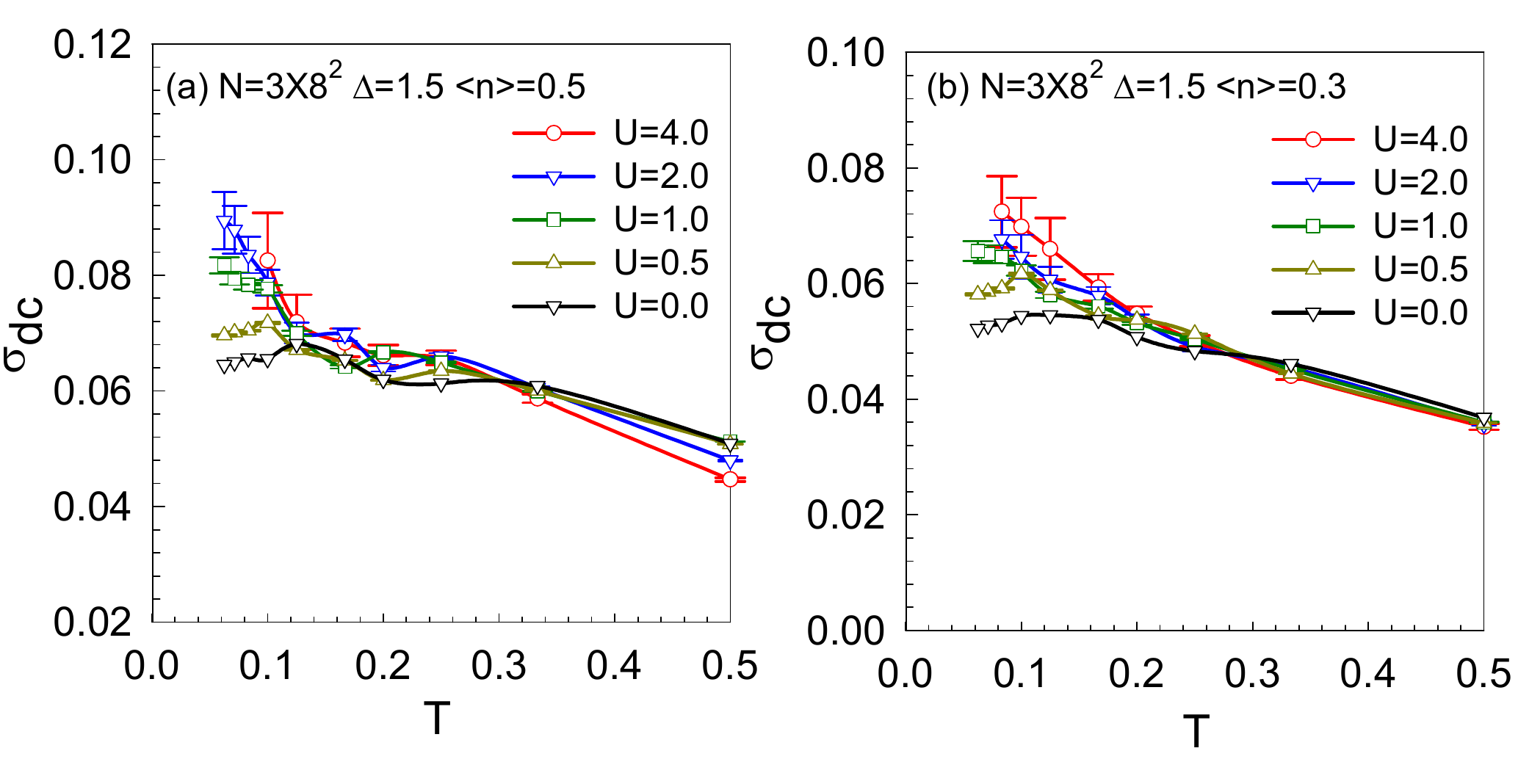}}
\caption{(Color online) The DC conductivity as a function of temperature for various disorder strengths at $\avg{n}=0.5$ and
$\avg{n}=0.3$.}
\label{Fig:U}
\end{figure}
In Fig.\ref{Fig:dc} (a), we show the behavior of the DC conductivity for a range of disorder
strengths at $\avg{n}=0.5$ and $U=4.0$. The DC conductivity curve below $\Delta=1.6$ is concave up ($d\sigma_{dc}/dT<0$) at low $T$,
indicating a metallic phase,
while above $\Delta=1.8$, the curve is concave down ($d\sigma_{dc}/dT>0$), corresponding to an insulating phase.
This figure displays a transition from metallic to insulating behavior at a critical disorder strength of $\Delta_{c}=1.6\sim1.8$,
completely differing from the half filling case (see Fig.\ref{Fig:half} (a)).
Similarly, $\Delta_{c}=1.8\sim2.0$ at $\avg{n}=0.3$ for $U=4.0$.
This indicates that disorder localizes the electrons.
We also focus on the effect of doping on transport properties.
For example, when keeping other parameters constant ($\Delta=1.4$ and $U=4$), the system at $\avg{n}=0.3,0.5$ is metallic, but is insulating at half filling; see Fig.\ref{Fig:half} (b)), suggesting that doping can lead to the metal-insulator transition.

The difference between the square lattice and Lieb lattice lies at the flat band; thus, one
can compare the physical properties of these two lattices to analyze the importance of the flat band.
At half filling, the Lieb lattice exhibits weaker metallicity than the square lattice with the same parameters; see Fig.\ref{Fig:half} (b) and Fig. 1 in Ref.\cite{PhysRevLett.83.4610}.
A similar phenomenon also appears at $\avg{n}$=0.3,0.5 when the flat band is empty.
Thus, the flat band will suppress transport properties regardless of whether the electrons have filled the flat band.

To observe the role of the interaction, $U$, on the DC conductivity, we compare the $\sigma_{dc}$ behavior of different interactions.
The value of disorder strength we choose is 1.5, so that the system size is larger than the localization length
and the disordered noninteracting system is insulating.
Fig.\ref{Fig:U} displays the temperature dependence of the DC conductivity for electronic
density $\avg{n}=0.5$ and $0.3$,
the interaction is found to play a profound role in the DC conductivity:
in the high-temperature regime, the interaction slightly reduces $\sigma_{dc}$;
In contrast, in the low-temperature regime, upon turning on the interaction,
the value of $\sigma_{dc}$ starts to increase,
and a noninteracting insulating phase gradually changes into a metallic phase at approximately $U=1.0$.
This phenomenon indicates that the interaction has a delocalizing effect on the electrons
in the disordered systems.

\begin{figure}[t]
\centerline {\includegraphics*[width=3.6in]{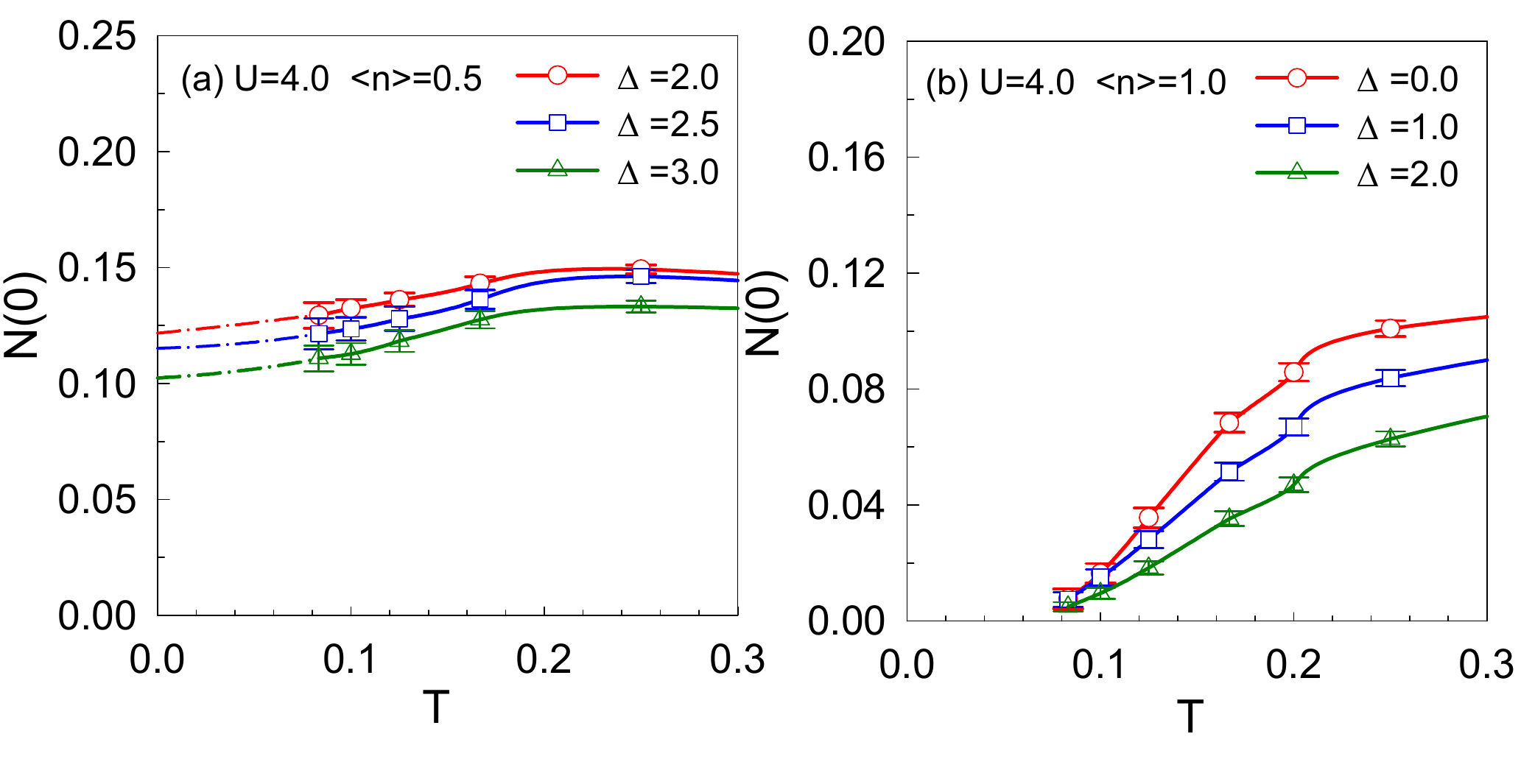}}
\caption{(Color online) The density of states as a function of temperature at (a) $\avg{n}=0.5$ and (b) $\avg{n}=1.0$ for $U=4.0$ for various disorder strengths.}
\label{Fig:dos}
\end{figure}
The single particle gap is another electronic property we are interested in. At half filing, the Hubbard model of the
the Lie lattice shows a charge excitation gap for $U>0$ without disorder, i.e., the properties of a Mott insulator.
Since the Anderson insulator is gapless at the Fermi level,
the gap can be used to determine the type of the insulating phase
although the gap is unaffected by  the order parameter of symmetry breaking.
Here, we use the density of states at the Fermi energy to extract the single-particle gap.
In Fig.\ref{Fig:dos} DOS is plotted as a function of temperature for various disorder strengths and fixed on-site Coulomb repulsive interaction
at different densities.
One can see that $N(0)$ decreases with decreasing temperature for all cases.
When $T\rightarrow0$, DOS will converge to  finite values at $\avg{n}=0.5$, suggesting the presence of the Anderson insulator. With increasing intensity of disorder, the
value of DOS decreases in the thermodynamic limit but is finite.
The finite density of states forms an interesting counterpoint to that of the Mott insulator whose DOS converges to zero, as shown in (b).

\begin{figure}[t]
\centerline {\includegraphics*[width=3.6in]{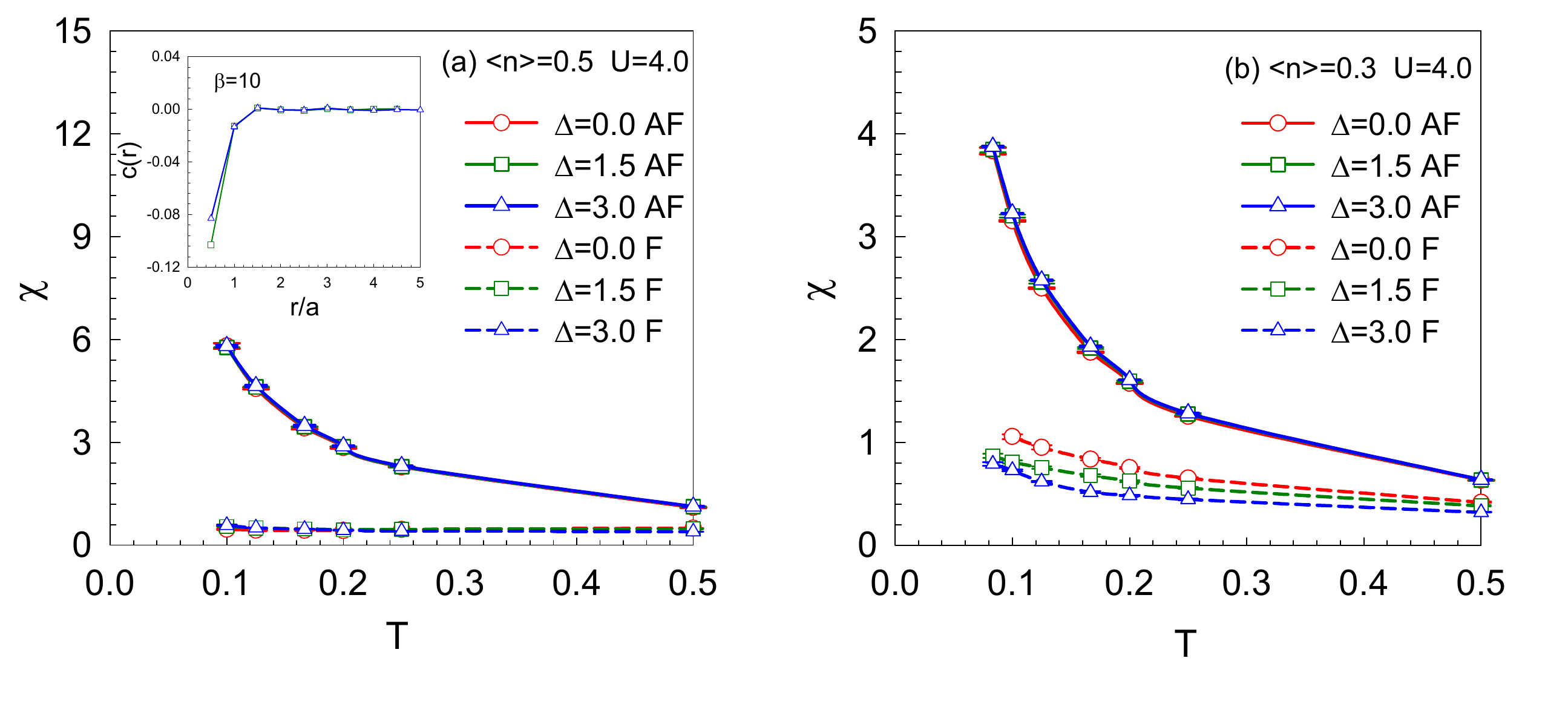}}
\caption{(Color online)(a) The ferromagnetic and anti-ferromagnetic susceptibility as a function of temperature at $\avg{n}=0.5,0.3$ for $U=4.0$. The inset:
spatial dependence of the spin-spin correlation function at $\beta=10$. $r$ running along a horizontal line, i.e., line with alternating red and blue sites in Fig.\ref{Fig1} (a).}
\label{Fig:sus}
\end{figure}
Finally, we consider the effect of doping on magnetic order.
Fig.\ref{Fig:sus} provides the dependence of susceptibility $\chi$ on $T$ for $U=4.0$ and various disorder strengths.
In all cases, it can be seen that the ferromagnetic susceptibility $\chi_{FM}$ decreases with decreasing temperature, while the
anti-ferromagnetic susceptibility $\chi_{AFM}$ increases and is always stronger than $\chi_{FM}$.
Although $\chi_{AFM}$ increases rapidly when $T$ approaches 0, it is still not enough to prove that there is an antiferromagnetic (AFM) order in our system. If there was perfect order, $\chi_{AFM}$ should approach $\beta\ast N_{c}$ when $T\rightarrow0$; however, $\chi_{AFM}$ here is much smaller than this value.
In the inset of Fig.\ref{Fig:sus}(a), inhibition of disorder is only apparent in non-nearest neighbor terms which contribute little to antiferromagnetism; perhaps this is why disorder has little effect on magnetic susceptibility.
Overall, Fig.\ref{Fig:sus} proves that the doped system is paramagnetic, while the half-filled system is ferromagnetic\cite{PhysRevB.94.155107}, indicating doping can change the magnetic properties.

\section{\label{intro}Conclusions}
In summary, we studied a doped Lieb lattice containing both
on-site Coulomb repulsion and disorder using a determinant quantum Monte Carlo method.
We computed the half-filled case to examine our method and explored the sign problem at states far away from half filling.
The sign problem becomes more serious with decreasing temperature and weaker with increasing disorder strength.
To ensure the reliability of the results, we chose two electron densities where the sign problem is less severe.
We calculated the temperature-dependent DC conductivity $\sigma_{dc}$ and found that
Hopping disorder drives the metallic phase to the insulating phase, while Coulomb repulsion has the opposite effect.
Compared to the previous theoretical results on the half-filled case, the produced insulating phase is proven to be an Anderson insulator by studying the density of states in the Fermi energy.
In addition, data on the susceptibility, $\chi(T)$, suggested that the doping system is in a paramagnetic state.
Compared to the ferrimagnetic half-filled system, it can be inferred that doping can change a lattice magnetic properties.

\begin{acknowledgments}
We thank Richard T. Scalettar and James E Gubernatis for many helpful discussions.
This work is supported by NSFC (No. 11974049). The numerical simulations were performed at the HSCC of Beijing Normal University and on the Tianhe-2JK in the Beijing Computational Science Research Center.
\end{acknowledgments}

\bibliography{reference}

\end{document}